\documentclass{PoS}
\usepackage{subfigure}
\usepackage{amsmath}
\usepackage{cite}

\title{The nonplanar cusp and collinear anomalous dimension at four loops in ${\mathcal N} = 4$ SYM theory}

\ShortTitle{The nonplanar cusp and collinear anomalous dimension at four loops in ${\mathcal N} = 4$ SYM}

\author{Rutger H. Boels\\
        II. Institut f\"ur Theoretische Physik, Universit\"at Hamburg,  \\ Luruper Chaussee 149, 22761 Hamburg, GERMANY\\
        E-mail: \email{Rutger.Boels@desy.de}}

\author{\speaker{Tobias Huber}%
	\\
       Naturwissenschaftlich-Technische Fakult\"at, Universit\"at Siegen, \\ Walter-Flex-Str.~3, 57068 Siegen, GERMANY\\
       E-mail: \email{huber@physik.uni-siegen.de}}

\author{Gang Yang\\
        CAS Key Laboratory of Theoretical Physics, Institute of Theoretical Physics, \\ Chinese Academy of Sciences, Beijing 100190, CHINA\\
        E-mail: \email{yangg@itp.ac.cn}}

\abstract{We present numerical results for the nonplanar lightlike cusp and collinear anomalous dimension at four loops in ${\mathcal N} = 4$ SYM theory, which we infer from a calculation of the Sudakov form factor. The latter is expressed as a rational linear combination of uniformly transcendental integrals for arbitrary colour factor. Numerical integration in the nonplanar sector reveals explicitly the breakdown of quadratic Casimir scaling at the four-loop order. A thorough analysis of the reported numerical uncertainties is carried out.}

\FullConference{13th International Symposium on Radiative Corrections (Applications of Quantum Field Theory to Phenomenology)\\
         25-29 September, 2017 \\
         St. Gilgen, Austria}

\begin{document}

\section{Introduction}
\label{sec:intro}

The lightlike cusp anomalous dimension (CAD) governs the structure of infrared (IR) divergences and as such is a universal quantity with many physical applications, see e.g.~\cite{Korchemsky:1985xj, Mueller:1979ih, Collins:1980ih, Sen:1981sd, Magnea:1990zb, Bern:2005iz}. The latter range from string theory~\cite{Gubser:2002tv} over collider physics (see e.g.~\cite{Moch:2004pa,Vogt:2004mw}) to flavour physics at scales as low as $m_b \sim 5$~GeV or even below~\cite{Bell:2010mg}. The lightlike CAD possesses a loop expansion, and in QCD its calculation was completed to three loops more than a decade ago~\cite{Moch:2004pa,Vogt:2004mw}.
Although QCD is the theory for phenomenology, there were tremendous achievements in $\mathcal{N}=4$ super Yang-Mills (SYM) theory in recent years by studying scattering amplitudes~\cite{N4review}, whose insights also catalyzed QCD calculations, e.g.\ for LHC processes. However, calculations in $\mathcal{N}=4$ SYM are usually restricted to the planar limit, in which the lightlike CAD was explicitly computed to four-loop accuracy~\cite{Bern:2006ew, Cachazo:2006az, Henn:2013wfa}, and predicted to all orders from integrability~\cite{Beisert:2006ez}. Beyond the planar limit, much less is known in general despite some very recent progress~\cite{Alday:2017xua,Aprile:2017bgs}. In the perturbative expansion of the CAD the first nonplanar correction enters at four loops, and no nonplanar correction had been computed in any theory until the first numerical four-loop result in $\mathcal{N}=4$ was presented in~\cite{Boels:2017skl}.

When analyzing the lightlike CAD in QCD one observes two interesting features. The first is the maximal transcendentality principle~\cite{Kotikov:2002ab,Kotikov:2004er}, a general conjecture that relates the maximal transcendental terms appearing in QCD directly to $\mathcal{N}=4$ SYM for certain quantities. It was verified to three loops in case of the lightlike CAD. The second feature concerns the fact that up to three loops the quark and gluon CAD are proportional to each other and differ only by the quadratic Casimir invariant of their respective gauge group representation. This property is known as \emph{quadratic Casimir scaling} of the CAD and plays an important role in IR factorization in gauge theories~\cite{Korchemsky:1988si, Gardi:2009qi, Dixon:2009gx, Becher:2009qa, Becher:2009kw, Dixon:2009ur, Ahrens:2012qz}. In~\cite{Becher:2009qa} it was even conjectured
that the nonplanar part of the CAD vanishes in any gauge theory, i.e.\  quadratic Casimir scaling should hold to all orders in perturbation theory. In~\cite{Frenkel:1984pz}, on the other hand, it was noted that quadratic Casimir scaling may be violated at higher orders of the perturbative expansion, see also \cite{Alday:2007mf}. Moreover, it is known that this scaling breaks down in $\mathcal{N}=4$ SYM at strong coupling~\cite{Armoni:2006ux}, and via instanton effects~\cite{Korchemsky:2017ttd}. Finally, our work~\cite{Boels:2017skl, Boels:2017ftb}, on which the present article is based on, disproved the conjecture in $\mathcal{N}=4$ SYM by an explicit computation of the four-loop nonplanar part of the lightlike cusp and collinear anomalous dimension via the Sudakov form factor. After~\cite{Boels:2017skl}, violation of Casimir scaling in QCD was reported in~\cite{Moch:2017uml,Grozin:2017css}.

\section{Form factor and anomalous dimensions}
\label{sec:ffad}

One of the simplest observables that contains the lightlike CAD is the Sudakov form factor. It can be obtained from the correlator of an operator of the stress-tensor multiplet with two on-shell massless states. Since the operator has zero anomalous dimension, the form factor has no ultraviolet but only IR divergences. Of particular interest are the loop expansion, the colour structure, and the structure of the IR divergences of the form factor, which we briefly review in turn in the following.

The perturbative expansion of the Sudakov form factor takes the form
\begin{equation}
{\cal F} = {\cal F}^{\textrm{tree}} \sum_{l=0}^\infty g^{2 {l}} (-q^2 )^{-l \epsilon} F^{(l)}  \, .
\end{equation}
If we denote the two on-shell momenta of the massless states by $p_1,p_2$, the off-shell quantity $q^2=(p_1+p_2)^2$ is the only dimensionful scale of the problem. In dimensional regularisation ($D=4-2 \epsilon$), $F^{(l)}$ is a function of gauge group invariants and $\epsilon$ only. The coupling constant $g$ reads $g^2 = \frac{g_{\rm YM}^2N_c}{(4\pi)^2}(4\pi e^{- \gamma_{\text{E}}})^\epsilon$. The Sudakov form factor in $\mathcal{N}=4$ SYM was computed to two loops in~\cite{vanNeerven:1985ja}. The three-loop correction was completed in~\cite{Gehrmann:2011xn} by using insights from the corresponding calculation in QCD~\cite{Baikov:2009bg,Lee:2010cga,Gehrmann:2010ue,Gehrmann:2010tu,vonManteuffel:2015gxa}. The integrand of the complete four-loop contribution was derived in~\cite{Boels:2012ew}, followed by its reduction to master integrals~\cite{Boels:2015yna}. Also the five-loop integrand is known~\cite{Yang:2016ear}.
 
Turning to the colour structure of the form factor, let us take $\mathcal{N}=4$ SYM with gauge group $SU(N_c)$ for definiteness here, although generalizations to other Lie groups are straightforward. The colour structure to $l \le 3$ loops is simply given by $(C_A)^l = N_c^l$. This changes starting from four loops since, besides $(C_A)^4 = N_c^4$, also the so-called quartic Casimir invariant $d_{A}^{abcd}d_{A}^{abcd}$ appears. The tensor $d_{A}^{abcd}$ is the symmetrized trace over four adjoint generators, $d_{A}^{abcd} = 1/6 \, {\rm Tr}[ T_A^a T_A^b T_A^c T_A^d + {\text{perms.}(b,c,d)} ]$,
with $[T_A^a ]_{xy} = -i f^{axy}$. For $SU(N_c)$ one has $d_{A}^{abcd}d_{A}^{abcd}/N_A = N_c^2/24 \, (N_c^2+36)$ and $N_A=(N_c^2-1)$.
Hence besides the {\emph{planar}} (i.e.\ $N_c^l$ leading-colour) contribution a {\emph{nonplanar}} (i.e.\ $N_c^{l-2}$ subleading-colour) correction enters at four loops. Starting from six loops, additional group invariants appear~\cite{Boels:2012ew}.

Finally, to see the connection between the form factor and the anomalous dimensions advocated in the introduction, one has to analyze the IR structure of the form factor.
Setting $q^2=-1$ and ${\cal F}^{\textrm{tree}}=1$, the IR structure takes the following form \cite{Bern:2005iz} (see also~\cite{Mueller:1979ih, Collins:1980ih, Sen:1981sd, Magnea:1990zb})
\begin{align}
\log {\cal F} = - \sum_{l=1}^\infty g^{2 l} \bigg[ \frac{\gamma_{\textrm{cusp}}^{({l})} }{(2 {l} \epsilon)^2} + \frac{{\cal G}_{\textrm{coll}}^{({l})} }{2 {l} \epsilon} + {\rm Fin}^{(l)} \bigg] + {\mathcal O}(\epsilon) \,, \label{eq:logFF}
\end{align}
where the leading and subleading singularity contains the cusp ($\gamma_{\textrm{cusp}}$) and collinear (${\cal G}_{\textrm{coll}}$) anomalous dimension, respectively.
The $l$-loop form factor $F^{(l)}$ has leading divergence $\propto 1/\epsilon^{2l}$. However, exponentiation of IR divergences ensures that in the logarithm of the form factor at most a double pole in $\epsilon$ remains. To this end, lower-loop contributions have to be expanded to higher terms in the Laurent expansion in $\epsilon$. For instance, the determination of ${\rm Fin}^{(l)}$ requires $F^{(1)}$ to ${\cal O}(\epsilon^{2l-2})$. As mentioned above, the first nonplanar correction starts at four loops, and the nonplanar part of the four-loop form factor takes the form
\begin{equation}\label{eq:centralrelation}
F^{(4)}_{\textrm{NP}}  = -\frac{\gamma_{\textrm{cusp, NP}}^{(4)}}{(8\epsilon)^2} - \frac{{\cal G}_{\textrm{coll,NP}}^{({4})} }{8 \epsilon} - {\rm Fin}_{\textrm{NP}}^{(4)} + {\mathcal O}(\epsilon) \,.
\end{equation}
In particular, it can at most have a double pole in $\epsilon$ instead of the full $1/\epsilon^8$ since, upon taking the logarithm in~(\ref{eq:logFF}), it cannot mix with any planar contribution from lower loops. We emphasize, however, that individual integrals that contribute to $F^{(4)}_{\textrm{NP}}$ will typically show the full $1/\epsilon^8$ divergence. The cancellation of these higher-order poles in the final result will provide a very strong constraint as well as an important consistency check of our computation.
The loop expansions of the cusp and collinear anomalous dimension read~\cite{Korchemskaya:1992je,Bern:2006ew,Cachazo:2006az,Gehrmann:2011xn,Henn:2013wfa}
\begin{align}
\gamma_{\textrm{cusp}} =&  8 g^2-16\zeta_2 g^4 -176\zeta_4 g^6 +(- 1752 \zeta_6 - 64\zeta_3^2 + \gamma_{\textrm{cusp, NP}}^{(4)}) g^8 + {\cal O}(g^{10})  \,, \label{eq:CAD} \\
{\cal G}_{\textrm{coll}} =& -4 \zeta^3 g^4 + (32 \zeta_5 + \frac{80}{3} \zeta_2\zeta_3) g^6 + ({\cal G}_{\textrm{coll, P}}^{(4)} + {\cal G}_{\textrm{coll, NP}}^{(4)}) g^8 + {\cal O}(g^{10})\, . \label{eq:collAD}
\end{align}
A numerical result for the planar part of the collinear anomalous dimension at four loops was given in~\cite{Cachazo:2007ad}. Here we will present numerical results for the nonplanar part of both the cusp and collinear anomalous dimension at four loops, which we will extract from the nonplanar part of the four-loop Sudakov form factor.

\section{Sudakov form factor in UT basis}
\label{sec:npffut}

The integrand of both the planar and nonplanar four-loop Sudakov form factor in $\mathcal{N}=4$ SYM was derived in~\cite{Boels:2012ew} based on colour-kinematics duality. There are $34$ integral topologies in total, each with $12$ propagators. $24$ ($14$) topologies contribute to the planar (nonplanar) part, and hence $4$ to both. The obtained integrals are very complicated for several reasons: Besides the number of loops, many topologies are crossed, and integrals typically have irreducible numerators of mass dimension four (i.e.\ two scalar products of loop and/or external momenta). In our present approach, we aim for a numerical evaluation of the integrals in a basis where each integral has so-called {\emph{uniform transcendentality (UT)}}, i.e.\ the constants that appear at a given order of the $\epsilon$-expansion have all the same overall transcendental weight, which increases by unit steps from one order in $\epsilon$ to the next. We therefore have to solve two problems. First, transform the four-loop integrand into a UT basis and second, integrate the appearing UT integrals numerically.

Starting with the first of these tasks, there are basically three ways to show that an integral is UT without explicitly computing it
\begin{itemize}
\item A UT integral can be written in the $d$Log form \cite{Arkani-Hamed:2014via, Bern:2014kca}.\\[-2.0em]
\item The leading singularities, or equivalently, the residues at all poles of a UT integral is always a constant \cite{Bern:2014kca,Bern:2015ple,Henn:2016men}.\\[-2.0em]
\item A UT integral basis leads to simple differential equations~\cite{Henn:2013pwa}.
\end{itemize}
Since the last item is subject to other articles in these proceedings~\cite{SteinhauserRADCOR}, we illustrate the first two UT properties here. For this purpose it is convenient to trade the four components of each loop momentum $l^{\mu}$ for four scalar parameters according to
\begin{equation}
l^{\mu} = \alpha_1 p_1^{\mu} + \alpha_2 p_2^{\mu} + \alpha_3 q_1^{\mu} + \alpha_4 q_2^{\mu} \, ,
\end{equation}
with $ p_i^2 = q_i^2 =  q_i \cdot p_j = 0 \quad \forall i,j$ and $q_1\cdot q_2 = - p_1\cdot p_2$.
In this so-called parametric form we now attempt another variable transformation such that the integral turns into the form $dw/w=d\log(w)$ for each integration variable. The latter form is then referred to as $d$Log form. If such a $d$Log form exists the integral is UT. We emphasize that finding a $d$Log form for generic four-loop form factor integrals is, in general, a difficult task. Moreover, the method of obtaining a $d$Log form is more suitable to {\emph{show}} the UT property of a given integral, rather than to {\emph{derive}} a UT numerator. Explicit examples of four-loop form factor integrals where a $d$Log form could be derived are given in~\cite{Boels:2017ftb}.

A procedure equivalent to deriving a $d$Log form is to study the leading singularity of an integral, which is most conveniently done in its parametric form. To this end one computes subsequent residues for all occurring scalar parameters ($16$ at four loops) in an initially chosen order. If, during this procedure, one encounters other than a simple pole in a remaining parameter, the integral is not UT. This method has the advantage that, besides checking the UT property of a single integral, it can be used to derive UT candidate integrals in a given topology. This derivation can be made algorithmic, and the algorithm constitutes an essential part towards the success of our calculation. For sake of brevity we sketch it here and refer to~\cite{Boels:2017ftb} for its full description. To determine UT numerators, we make an ansatz of mass dimension four which is a linear combination of all products of scalar products.
The requirement of avoiding other than simple poles in the procedure of taking residues gives constraints on the coefficients in the ansatz 
and finally enables us to obtain a set of UT candidate integrals in each topology. In practice we take a few hundred to thousand random residue orders to overcome the large ($16! \sim 2\times 10^{13}$) total number of different orders in which the residues can be taken.

\begin{figure*}[t]
\includegraphics[width=0.99\textwidth]{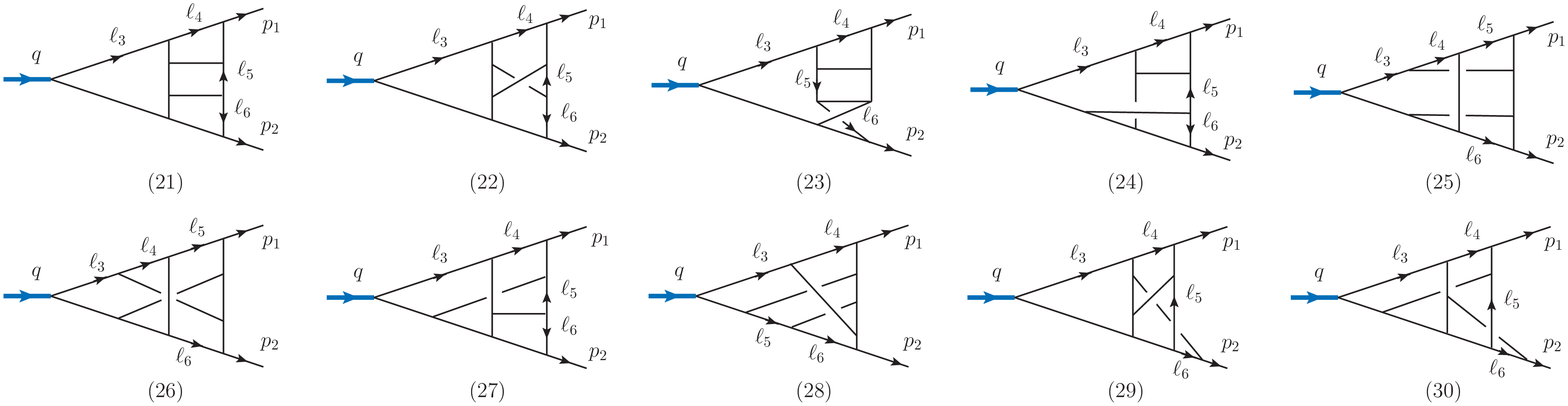}
\caption{\label{fig:NPtops}Topologies that contribute to the non-planar form factor at four loops in a basis of UT integrals.}
\end{figure*}

Having the UT candidates at hand we perform at least $10^5$ additional residue checks and/or derive $d$Log forms. Afterwards, we write the entire four-loop form factor in $\mathcal{N}=4$ SYM as a rational linear combination of UT integrals by means of the relations from the IBP reduction~\cite{Boels:2015yna,Boels:2016bdu}. While the full four-loop form factor in the UT basis can be found in~\cite{Boels:2017ftb}, we restrict the current presentation to the nonplanar part. It turns out that only $10$ out of $14$ integral topologies contribute, the remaining $4$ (labelled $(31)$~--~$(34)$ in~\cite{Boels:2012ew}) do not contain any UT integrals. Below we list the $23$ UT integrals $I^{(n)}_{1\,-\,23}$ that appear in the nonplanar form factor at four loops. The superscript $(n)$ denotes the twelve propagators from topology $(n)$ in Fig.~\ref{fig:NPtops}, such that we only have to list the numerator of each integral. The nonplanar form factor is then obtained as
\begin{equation}
\label{eq:FFNP}
F^{(4)}_{\textrm{NP}} = {48 \over N_c^2} \sum_{i=1}^{23} \, c_i \, I^{(n_i)}_{i} \,,
\end{equation}
with \, $\vec c = \{1/2, 1/2, 1/2, -1, 1/4, -1/4, -1/4, 2, 1, 4, 1, 1, -1/2, 1, 1, 1, 1, 1, 1, 1, -1, 1/4, 1/2\}$. \\
The prefactor $48/N_c^2 = 2 \times 24/N_c^2$ in~(\ref{eq:FFNP}) has its origin in the permutational sum over external legs and the colour factor~\cite{Boels:2012ew}. The UT integrals are 
\small
\allowdisplaybreaks
\begin{align}
I^{(21)}_{1}   &=[(\ell_3-p_1)^2]^2\\[0.0em]
I^{(22)}_{2}   &=(\ell_3-p_1)^2 \, [\ell_4^2+\ell_6^2-\ell_3^2+(\ell_3-\ell_4+p_1)^2  +(\ell_3-\ell_6-p_1)^2]\\[0.0em]
I^{(23)}_{3}   &=[(\ell_3-p_1)^2]^2\\[0.0em]
I^{(24)}_{4}   &=(\ell_3-p_1)^2 \, [(q-\ell_3-\ell_5)^2 + (\ell_5+p_2)^2 ]\\[0.0em]
I^{(25)}_{5}   &=\left[(p_1 - \ell_5)^2+2 (\ell_4-\ell_5)^2+(\ell_3-\ell_4)^2-(\ell_3-\ell_5)^2 -(p_1-\ell_4)^2 \right]^2 \nonumber \\[0.0em]
               & \quad  -4\, (\ell_4-\ell_5)^2 \, (p_1-\ell_3+\ell_4-\ell_5)^2 \\[0.0em]
I^{(26)}_{6}   &=[(\ell_3-\ell_4-\ell_5)^2-(\ell_3-\ell_4-p_1)^2-(\ell_6-p_2)^2-\ell_5^2]  [\ell_5^2-\ell_4^2-\ell_6^2+(\ell_4-\ell_6)^2] \nonumber \\[0.0em] 
               & \quad  +4\, \ell_5^2 \, (\ell_6-p_2)^2 + (\ell_4-\ell_5)^2 \, (\ell_3-\ell_4+\ell_6-p_2)^2 \\[0.0em]
I^{(26)}_{7}   &=4\, [(\ell_4-\ell_5) (\ell_3-\ell_4+\ell_5-p_1)]   [(\ell_4-\ell_6) (\ell_3-\ell_4+\ell_6-p_2)] \nonumber \\[0.0em]
	       & \quad - 4\, (\ell_4-\ell_5)^2 \, (\ell_3-\ell_4+\ell_6-p_2)^2 - (\ell_3-\ell_4)^2 \, (\ell_5+\ell_6-\ell_4)^2 \nonumber \\[0.0em]
               &  \quad  - \ell_6^2 \, (\ell_5-p_1)^2 - \ell_5^2 \, (\ell_6-p_2)^2  - \ell_4^2 \, (\ell_3-\ell_4+\ell_5+\ell_6-q)^2 \\[0.0em]
I^{(27)}_{8}   &=\frac{1}{2} \left[\ell_3^2 - \ell_4^2 - (\ell_4-\ell_3-p_1)^2\right]   \left[(\ell_3-\ell_4-\ell_5)^2 + (\ell_5+p_2)^2\right] \\[0.0em]
I^{(28)}_{9}   &=(\ell_3 - \ell_4 - p_2)^2 \, \left[ (\ell_3-\ell_4)^2 - (\ell_3-p_1)^2\right]\\[0.0em]
I^{(29)}_{10}  &=\frac{1}{2} \left[\ell_3^2 - \ell_4^2 - (\ell_4-\ell_3-p_1)^2\right] \, \left[\ell_6 \cdot (\ell_6 - \ell_4 + \ell_3 - p_2)\right]\\[0.0em]
I^{(30)}_{11}  &= (\ell_3-\ell_4-p_2)^2 [(p_1-\ell_4)^2+(\ell_3-\ell_4)^2-(\ell_3-p_1)^2] \\[0.0em]
I^{(27)}_{12}  &= \frac{1}{2} \, (\ell_3-\ell_4)^2 \, \left[2 \, (\ell_4-p_2)^2 + (\ell_6-p_1)^2 - \ell_4^2 + \ell_5^2 - (\ell_4-\ell_6)^2 + 2 \, (p_1+p_2)^2\right] \\[0.0em]
I^{(28)}_{13}  &= \frac{1}{2} \, (\ell_3-\ell_4)^2 \, \left[2 \, (\ell_3-\ell_4-p_2)^2 + (\ell_6-p_1)^2 + \ell_4^2   - (\ell_4-\ell_6)^2 \right] \\[0.0em]
I^{(29)}_{14}  &= (\ell_4-p_1)^2 \, \left[(\ell_3-\ell_4+\ell_6)^2 + (\ell_6-p_2)^2 - \ell_6^2\right] \\[0.0em]
I^{(29)}_{15}  &= \frac{1}{2} \, (\ell_3-p_1-p_2)^2 \, \left[(\ell_4-\ell_6)^2 - (\ell_4-p_2)^2  - (\ell_6-p_1)^2 - (p_1+p_2)^2 \right]\\[0.0em]
I^{(30)}_{16}  &= (\ell_3-p_1-p_2)^2 \, (\ell_5+p_2)^2 \\[0.0em]
I^{(30)}_{17}  &= \frac{1}{2} \, (\ell_4-p_1)^2 \, \left[2\, (\ell_5+p_2)^2 - (\ell_5+p_2+\ell_4-\ell_3)^2 \right] \\[0.0em]
I^{(30)}_{18}  &= \frac{1}{2} \, (\ell_3-\ell_4)^2 \, \left[2\, (\ell_6-\ell_4+p_1)^2 - 3 \, \ell_6^2 \right] \\[0.0em]
I^{(22)}_{19}  &= (\ell_3-\ell_4)^2 \,  (p_1-\ell_3+\ell_6)^2 \\[0.0em]
I^{(22)}_{20}  &= \ell_6^2 \,  (p_1-\ell_4)^2 \\[0.0em]
I^{(24)}_{21}  &= (p_1 - \ell_3-\ell_5)^2 \,  (\ell_3-p_1-p_2)^2 \\[0.0em]
I^{(24)}_{22}  &= \ell_5^2 \,  (\ell_3-p_1-p_2)^2 \\[0.0em]
I^{(28)}_{23}  &= (\ell_4 - p_1)^2 \,  (\ell_3-\ell_4+\ell_5-p_2)^2 \, . 
\end{align}
\normalsize

\section{Numerical integration and error analysis}
\label{sec:numinterror}

In order to numerically integrate the $23$ UT integrals in the nonplanar sector of the four-loop form factor we choose two main strategies: sector decomposition and Mellin-Barnes (MB) techniques. Within the sector decomposition approach two computer implementations have been used: mostly FIESTA~4~\cite{Smirnov:2015mct}, with cross-checks for simpler integrals using SecDec~3~\cite{Borowka:2015mxa}. The numerical integration with FIESTA is done with the VEGAS algorithm \cite{Lepage:1980dq} from the CUBA library \cite{Hahn:2004fe}. In SecDec, the CUHRE and DIVONNE algorithms are applied. Let us mention here the important empirical observation that in sector decomposition UT integrals usually generate considerably fewer integration terms compared to non-UT siblings of comparable complexity. Despite this simplification, four-loop form factor integrals remain challenging to integrate owing to the appearance of IR divergences. Typical runtimes are of the order of several weeks.

A second technique applied here is MB integration. Besides using automated tools such as MB~\cite{Czakon:2005rk,Smirnov:2009up} and AMBRE~\cite{Gluza:2007rt, Gluza:2010rn, Blumlein:2014maa} we have to overcome the problem that it is not straightforward to derive valid MB representations for crossed four-loop topologies with the loop-by-loop approach.
For this purpose we constructed an in-house MATHEMATICA routine based on a hybrid of the loop-by-loop approach and using the ${\cal F}$ and ${\cal U}$ graph polynomials.
Although this in principle renders valid MB representations for all crossed four-loop topologies, the obtained representations are in many cases too high-dimensional to be integrated in practice. Still, efficient MB representations could be found for some (planar and crossed) integrals, in which cases the precision of the numerical integration is typically 3 to 4 orders of magnitude better compared to sector decomposition. Moreover, the runtimes are much shorter, typically a few days. More details and examples on both sector decomposition and MB techniques are given in~\cite{Boels:2017ftb}.

\begin{figure}[t]
  \subfigure[]{
    \label{fig:subfig:a} 
    \hspace*{-6pt}\includegraphics[width=7.2cm]{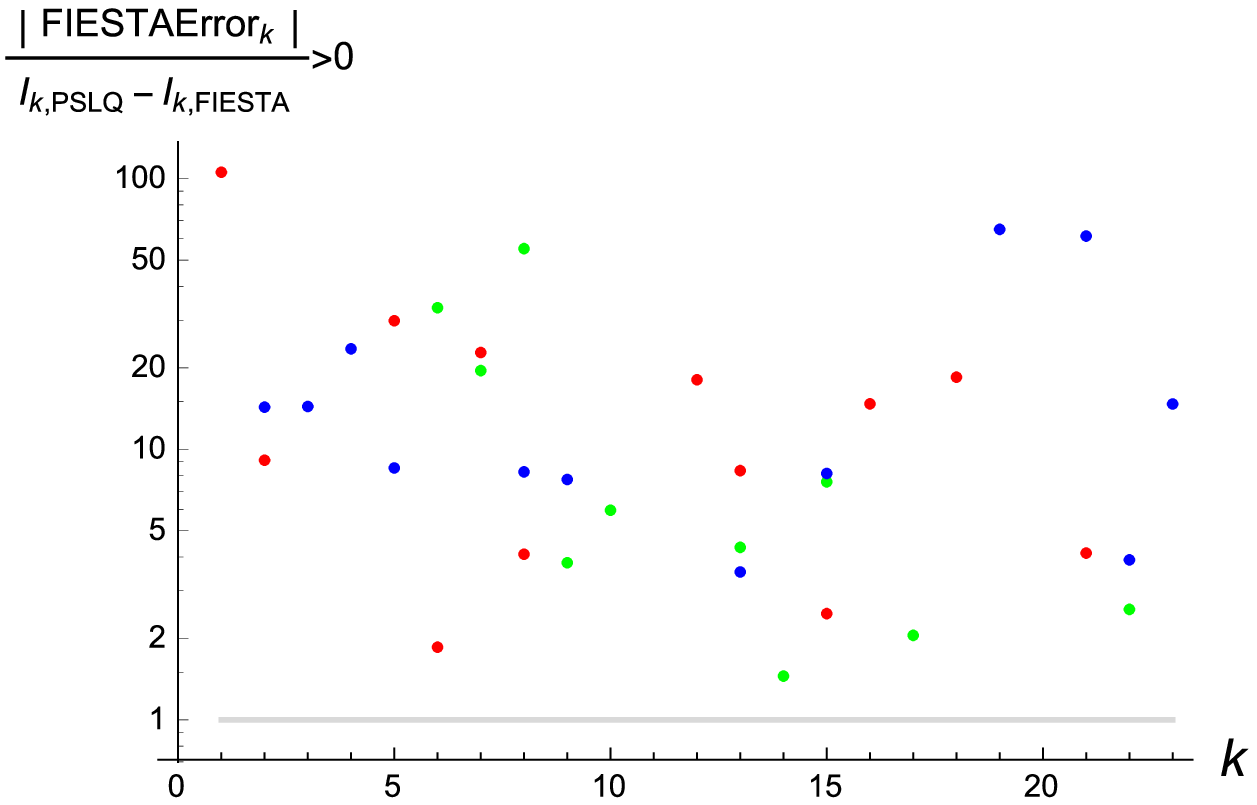}}
  \subfigure[]{
    \label{fig:subfig:b} 
    \hspace*{-10pt}\includegraphics[width=8.2cm]{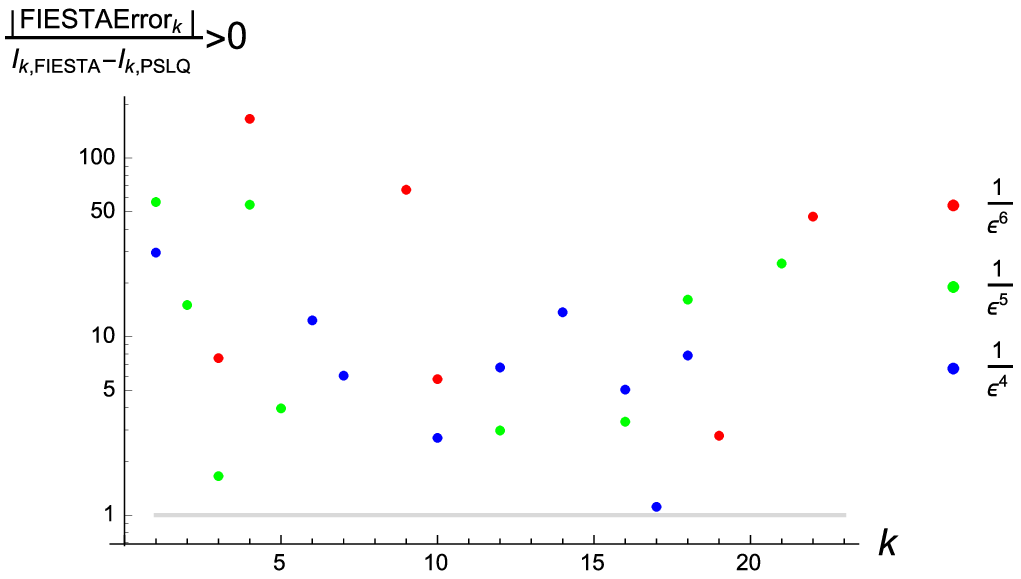}}
  \caption{Scatterplot of the relative error of FIESTA results compared to analytic (PSLQ) results for $\epsilon^{\{-6,-5,-4\}}$ orders. (a) Plot of cases ${\textrm{FIESTA error} \over I_{\rm PSLQ} - I_{\rm FIESTA}} > 0$. (b)  Plot of cases ${\textrm{FIESTA error} \over I_{\rm FIESTA} - I_{\rm PSLQ}} > 0$. All ratios larger than $200$ are not shown. All ratios are larger than unity.}
  \label{fig:PSLQcheck} 
\end{figure}

Since we use numerical integration methods, a thorough discussion of the errors in these integrals is needed before we can present our results. In as many cases as possible we performed the integration using two independent methods (Sector decomposition, MB techniques and an analytic result from~\cite{Henn:2016men}). However, most of the coefficients needed for the cusp and collinear anomalous dimension at order $\epsilon^{\{-2,-1\}}$ were obtained using exclusively the VEGAS algorithm in FIESTA. To check that the Gaussian regime in VEGAS is reached, we evaluate each integral for several evaluation point settings, and make sure the error scales as $1/\sqrt{\textrm{eval points}}$. For all integrals in the nonplanar sector of the form factor, this was reached very quickly. We have also checked that fluctuations upon increasing the number of sampling points are well within the reported error bars of previous runs where fewer points were used.

Luckily, the precision of our numerical result allows for another non-trivial cross-check: For the poles at orders $\epsilon^{\{-8,-6,-5,-4\}}$ the precision of the numerical integration is good enough to allow for a conversion of the reported central values into small rational multiples of $\{1,\zeta_2,\zeta_3,\zeta_4\}$ in the spirit of PSLQ~\cite{Ferguson:1999:API:307090.307114}. This can be used to obtain an estimate of the true precision for about seventy data points.
Then, we compute the ratio between FIESTA errors and the assumed 'true' errors obtained by comparing the reported central value to the PSLQ result at orders $\epsilon^{\{-6,-5,-4\}}$ via,
\begin{equation}
{\textrm{FIESTA error}_k \over I_{k,{\rm PSLQ}} - I_{k,{\rm FIESTA}}} \,,
\end{equation}
where $k$ labels the $23$ integrals from section~\ref{sec:npffut}. The results are depicted in figure~\ref{fig:PSLQcheck}, which contains two panels for positive and negative deviations, respectively.
We emphasize that for all $23$ integrals, all ratios have absolute values larger than unity, which indicates that the reported FIESTA errors are always larger than the discrepancy between the PSLQ result and the central value from numerical integration.
Moreover, by comparing the two panels in figure~\ref{fig:PSLQcheck}, it is clear that there is no definite sign of the deviation. The presence of a definite sign might have indicated a systematic error.

In total, our error analysis shows that the uncertainties reported by FIESTA are stable and in general conservatively estimate the errors for the form factor integrals in the present study. We will therefore interpret the FIESTA reported error as statistical, and as representing the standard deviation of a Gaussian error. As a consequence, the obtained individual errors are added in quadrature to obtain the total error. For reference, also the result of adding errors linearly is provided, although there is no hint for a systematic error in the case at hand.

\section{Results and conclusion}
\label{sec:results}

\begin{table*}[t]
\caption[]{Nonplanar form factor result and errors. The prefactor ${48 /N_c^2}$ in (\ref{eq:FFNP}) is not included.}
\label{tab:error}
\centering
\vspace*{5pt}

\resizebox{0.8\textwidth}{!}{
\begin{tabular}{l | c| c| c | c } 
\hline\hline
$\epsilon$ order & \, $-$8 \, & \, $-$7 \,& \, $-$6 \,& \, $-$5 \,    \cr \hline 
result           & $-3.8\times 10^{-8}$ & $+4.4\times 10^{-9}$ & $-1.2\times 10^{-6}$ & $-1.2\times 10^{-5}$ \cr \hline 
uncertainty      & -- & $\pm 5.7\times 10^{-7}$ & $\pm 1.0\times 10^{-5}$ & $\pm 1.2\times 10^{-4}$ \cr \hline \hline
\end{tabular}}

\vspace*{8pt}

\resizebox{0.62\textwidth}{!}{
\begin{tabular}{l | c| c| c | c} 
\hline\hline
$\epsilon$ order & \, $-$4 \,  & \, $-$3 \, & \, $-$2 \, & \, $-$1 \,    \cr \hline 
result   	 &  $+3.5\times 10^{-6}$ & $+$~0.0007 & $+$1.60 & \,$-$17.98 \,   \cr \hline 
uncertainty    	 &  $\pm 1.5\times 10^{-3}$  & $\pm$ 0.0186  & $\pm$0.19 & $\pm$ 3.25 \cr \hline \hline
\end{tabular}}

\end{table*}

Adding up the results for the integrals in the nonplanar part of the four-loop Sudakov form factor as described in section~\ref{sec:npffut} yields the numbers as shown in table~\ref{tab:error}. Each column contains the central value and the total uncertainty at a given order in the $\epsilon$-expansion. According to our reasoning in section~\ref{sec:numinterror} the total uncertainty was obtained by adding individual ones in quadrature. 

On physical grounds (eq.~(\ref{eq:centralrelation})), the coefficients at orders $\epsilon^{\{-8,-7,-6,-5,-4,-3\}}$ must vanish in the final result, which is indeed well satisfied within error bars. This provides a strong consistency check of our calculation. Whereas the coefficients at order $\epsilon^{-7}$ must vanish in each of the $23$ UT integrals $I^{(n)}_{1\,-\,23}$ separately, the orders $\epsilon^{\{-8,-6,-5,-4,-3\}}$ are non-zero in most of the $I^{(n)}_{1\,-\,23}$ (see appendix~A of~\cite{Boels:2017ftb}) but cancel in the final result. As mentioned earlier, the precision of the orders $\epsilon^{\{-8,-6,-5,-4\}}$ is good enough to allow for a conversion of the reported numbers \`a la PSLQ~\cite{Ferguson:1999:API:307090.307114} into small rational multiples of $\{1,\zeta_2,\zeta_3,\zeta_4\}$. After this conversion, these orders cancel even analytically in the final result of the nonplanar form factor.

Of particular interest is now the coefficient at order ${\cal O}(\epsilon^{-2})$, since it is directly related to the nonplanar four-loop CAD. Our result $1.60 \pm 0.19$ (see table~\ref{tab:error}) is clearly non-zero, with a statistical significance of~$8.4 \sigma$. Adding individual uncertainties linearly to account for potential systematic effects yields $1.60 \pm 0.58$, which is still significantly non-zero. Taking all relevant prefactors from eqs.~(\ref{eq:centralrelation}) and~(\ref{eq:FFNP}) into account allows us to translate this into a number for the nonplanar four-loop CAD. For gauge group $SU(N_c)$ we get
\begin{equation}
\label{eq:CAD-result}
\gamma_{\textrm{cusp, NP}}^{(4)}  = -3072\times( 1.60 \pm 0.19 ) \, \frac{1}{N_c^2} \, .
\end{equation}
It has the same sign as the planar result~\cite{Bern:2006ew, Beisert:2006ez, Cachazo:2006az, Henn:2013wfa} $\gamma_{\rm cusp, P}^{(4)} = - 1752 \zeta_6 - 64\zeta_3^2\sim-1875$. If $N_c=3$ is used, its value becomes $\gamma_{\textrm{cusp, NP}}^{(4)} \sim - 546 \pm 65$ and hence a factor of 3~--~4 smaller than the planar contribution. We conclude from this that quadratic Casimir scaling is violated starting from the four-loop level.

From the result at order ${\cal O}(\epsilon^{-1})$ the nonplanar collinear anomalous dimension (AD) at four loops can be derived. Using the result from table~\ref{tab:error} gives
\begin{equation}
\label{eq:CollAD-result}
{\cal G}_{\textrm{coll, NP}}^{(4)}  = -384\times( -17.98 \pm 3.25 ) \, \frac{1}{N_c^2} \, .
\end{equation}
There are two interesting features about this result. First, there is clear evidence that this result is non-zero as well, being in tension with a vanishing result at the ~$5.5\sigma$ level.  We mention that the linearly summed error is obtained as $-17.98 \pm 11.89$.
Second, the central value has opposite sign compared to the four-loop planar collinear AD result~\cite{Cachazo:2007ad}, ${\cal G}_{\textrm{coll, P}}^{(4)} = - 1240.9(3)$; and its sign is also different from that of the nonplanar cusp AD above. However, one has to keep in mind here that, unlike the cusp AD, the collinear AD is a scheme-dependent quantity. It would certainly be interesting to confirm the sign of the collinear AD by other methods.

To conclude, we achieved to express the entire Sudakov form factor in ${\mathcal N} = 4$ SYM theory as a rational linear combination of a few dozens of UT integrals only, which confirms the remarkable simplicity of this quantity already observed at lower loop orders and, moreover, demonstrates the power of techniques related to uniformly transcendental bases in maximal super-Yang Mills theory. Moreover, we arrived at the numerical integration of the nonplanar part of the form factor through to order ${\cal O}(\epsilon^{-1})$. Our numerical results show explicitly the breakdown of quadratic Casimir scaling from the first possible (i.e.\ four-loop) order. Moreover, the four-loop collinear AD seems to have opposite sign compared to its planar counterpart.

The crucial steps towards these results were on the one hand the development of an algorithm -- partially based on the ideas in~\cite{Arkani-Hamed:2014via,Bern:2014kca,Bern:2015ple,Henn:2016men} -- to construct UT candidate integrals from the principle of constant leading singularities. On the other hand, numerical integration routines based on sector decomposition and Mellin-Barnes techniques, together with large computing resources, were essential for stable numerical results and for reported error bars small enough to get a conclusive picture of the physical quantities that govern the structure of infrared divergences.

Four-loop calculations in gauge theories will remain a vivid topic also in the future. The computations of the four-loop quark and gluon form factors in QCD are in progress~\cite{Henn:2016men, Lee:2016ixa, vonManteuffel:2016xki, Lee:2017mip}. Interesting further directions also involve the computation of the angle-dependent nonplanar cusp AD in ${\mathcal N} = 4$ SYM theory at four loops, or the extension of the present calculation to five loops. The relevant integrand is already known~\cite{Yang:2016ear}. Moreover, it would be interesting to investigate the consequences of the breakdown of quadratic Casimir scaling to factorization theorems formulated in QCD and soft-collinear effective theory.

\section*{Acknowledgments}

TH would like to thank the organisers of ``RADCOR 2017'' for creating a very pleasant and inspiring atmosphere.
The work of RB is supported by the DFG within the SFB 676 ``Particles, Strings and the Early Universe''. The work of GY is supported in part by the CAS Hundred-Talent Program, by the Key Research Program of Frontier Sciences of CAS, and by Project 11647601 supported by National Natural Science Foundation of China.

\providecommand{\href}[2]{#2}\begingroup\raggedright\endgroup

\end{document}